\documentclass{emulateapj}
\usepackage{apjfonts}
\usepackage{natbib}
\usepackage{amsmath}
\usepackage{graphicx}
\def \ni {\noindent}
\def\lsim{\mathrel{\rlap{\lower4pt\hbox{\hskip1pt$\sim$}}
    \raise1pt\hbox{$<$}}}                
\def\gsim{\mathrel{\rlap{\lower4pt\hbox{\hskip1pt$\sim$}}
    \raise1pt\hbox{$>$}}}                

\def\kms{$\text{km s}^{-1}$}
\def\nbin{$N_{\mathrm{bin}}$}

\def\apjl{ApJL}
\def\aj{AJ}
\def\mnras{MNRAS}
\def\apj{ApJ}
\def\araa{ARA\&A}
\def\nat{Nature}
\def\aap{A\&A}
\def\apjs{ApJS}

\shorttitle{Non-parametric Schwarzschild Models of Draco}

\defcitealias{nav96}{NFW}

\begin{document}

\slugcomment{{\sc Accepted to ApJ:} 11 November 2012} 

\title{Measuring Dark Matter Profiles Non-Parametrically in Dwarf Spheroidals:\\
An Application to Draco}

\author{John R. Jardel\altaffilmark{1},
  Karl Gebhardt\altaffilmark{1},
  Maximilian Fabricius\altaffilmark{2},
  Niv Drory\altaffilmark{3},
  and Michael J. Williams\altaffilmark{2}}

\altaffiltext{1}
{The University of Texas, Department of Astronomy,
2515 Speedway, Stop C1400, Austin, Texas 78712-1205;\\
jardel@astro.as.utexas.edu}
\altaffiltext{2}{Max-Planck Institut f\"{u}r extraterrestrische Physik, 
Giessenbachstrasse, D-85741 Garching bei M\"{u}nchen, Germany}
\altaffiltext{3}{Instituto de Astronom\'{i}a,
Universidad Nacional Aut\'{o}noma de M\'{e}xico,
Avenida Universidad 3000, Ciudad Universitaria, C.P. 04510
M\'{e}xico D.F., M\'{e}xico}

\begin{abstract}

We introduce a novel implementation of orbit-based (or Schwarzschild) modeling
that allows dark matter density profiles to be calculated non-parametrically
in nearby galaxies.  Our models require no assumptions to be made about 
velocity anisotropy or the dark matter profile.  The technique can be applied
to any
dispersion-supported stellar system, and we demonstrate its use by studying
the Local Group dwarf spheroidal (dSph) galaxy Draco.  We use existing 
kinematic data at larger radii and also present 12 new radial velocities within
the central $13$~pc obtained with the VIRUS-W 
integral field spectrograph on the 2.7m telescope at McDonald Observatory.
Our non-parametric Schwarzschild models 
find strong evidence that the dark matter profile in Draco is cuspy for
$20 \leq r \leq 700$~pc.  The profile for $r \geq 20$~pc is well-fit by a 
power law with slope $\alpha = -1.0 \pm 0.2$, consistent with predictions
from Cold Dark Matter (CDM) simulations.  Our models confirm that, despite its
low baryon content relative to other dSphs, Draco lives in a massive halo.

\end{abstract}


\section{Introduction}

Understanding how dark matter is distributed in low-mass galaxies is central
to the study of galaxy formation in the cold dark matter (CDM) paradigm.
The first CDM simulations predicted that
all dark matter halos share a universal density profile with a cuspy inner slope
of $\alpha \equiv d \ln \rho_{DM}/ d \ln r = -1$ (\citealt{nav96}, hereafter 
\citetalias{nav96}).  When observers began studying low-mass galaxies, however,
they mostly found halos with a uniform density $\alpha=0$ core 
\citep{bur95,per96,bor01,deb01,bla01,sim05}. 
This disagreement between
theorists and observers over the form of $\rho_{DM} (r)$
became known as the core/cusp debate.

Since the debate began, the number of profile parameterizations used to describe
low-mass galaxies by both
theorists and observers has only increased.
Observers champion empirical fits such as the Burkert profile
\citep{bur95,sal00}, cored isothermal models \citep{per96}
or simply generic broken power laws \citep{koc07,str08,wal09b}.
Theorists have also introduced new fits to their simulated halos with varying,
although still cuspy, inner slopes \citep{nav04,sta09,nav10}.  Modeling a
galaxy with each of these parameterizations would not only be time 
consuming, but also biased if the true profile is unlike any of them.  It is 
therefore preferable to use non-parametric methods to determine $\rho_{DM}(r)$.

Van den Bosch et al. (2006) first experimented with
non-parametric orbit-based models by allowing the mass-to-light ratio $M/L$
to vary with radius in the globular cluster M15.
We introduce a similar modeling technique that 
uses axisymmetric Schwarzschild modeling, combined with knowledge of the 
full line-of-sight
velocity distribution (LOSVD) of stars, to break the well-known degeneracy 
between mass and orbital anisotropy.  We demonstrate the capability of these 
models by applying them to the Local Group dwarf spheroidal (dSph) galaxy
Draco.  Draco is part of an interesting class of objects that are some of the
most dark matter-dominated galaxies discovered.  This makes differentiating 
between dark and luminous mass in dSphs easier as the baryons have less of an 
effect on the total density profile than they do in larger galaxies.
Recently, using improved data and modeling techniques, \citet{ada12} 
found a cuspy dark matter profile in the low-mass galaxy NGC~2796 where 
previous studies found a core.  Studies like these motivate us to 
investigate the dSphs with more sophisticated models.

Our models represent a significant improvement over previous work on dSphs
as most studies use spherical Jeans models \citep{gil07,wal09b,wol10} 
which require the modeler to
make assumptions about the nature and degree of the anisotropy.  These 
assumptions vary in complexity from simply assuming isotropy, which can bias
the estimate of $\alpha$ \citep{eva09},
to parameterizing the anisotropy as a general function of radius 
\citep{str08,wol10}.  
Models that allow for more freedom in the anisotropy typically
fall victim to the mass-anisotropy degeneracy and cannot robustly determine
the inner slope of $\rho_{DM}(r)$ \citep{wal09b}.  We hope to make a robust
determination of the inner slope in Draco with a suite of more general 
non-parametric Schwarzschild models.

\section{Non-parametric Schwarzschild Models}

At the heart of our non-parametric technique is the orbit-based modeling
code developed by \citet{geb00b,geb03}, updated by \citet{tho04,tho05},
and described in detail in \citet{sio09}.  All orbit-based codes are based
on the principle of orbit superposition first introduced by \citet{sch79}.
Similar axisymmetric codes are used by
\citet{rix97},\citet{vdm98}, \citet{cre99}, and \citet{val04} while 
\citet{vdb08} present a 
fully triaxial modeling code.  The current Schwarzschild models that allow 
for dark matter do so by requiring the modeler to assume a parameterization
for the dark matter density profile $\rho_{DM}(r)$.  Unfortunately, this
parameterization is often exactly what we wish to determine.  Current methods
get around the circular logic of this dilemma by running models with different 
parameterizations and comparing their relative goodness-of-fit with a $\chi^2$
test.  Non-parametric modeling sidesteps the issue entirely, and lets the 
parameterization of $\rho_{DM}(r)$ be output from the models, rather than
input as a guess.  

The principle of orbit superposition works by choosing from a library of all 
possible stellar orbits only those orbits that best reproduce the observed
kinematics of the galaxy being modeled.  If we know the mass density profile
of the galaxy, and hence the potential, we can compute the appropriate orbit
library.  However, since we do not know the potential of the galaxy, we must
construct a number of models with slightly different mass distributions and
compare the goodness-of-fit of the resulting allowed orbits.  The
radial profile of the total (dark + stellar) mass density in a galaxy can
be written as:

\begin{equation}
\rho(r) = \frac{M_*}{L} \times \nu(r) + \rho_{DM}(r)
\label{param}
\end{equation}

\ni
where $M_*/L$ is the mass-to-light ratio of the stars, $\nu(r)$ is the stellar
luminosity density profile, and $\rho_{DM}(r)$ is the dark matter density
profile.  In principle we know $M_*/L$, which can vary as a function of radius,
from stellar population models.  We also know $\nu(r)$ from the de-projection 
of the observed surface brightness profile.  Our task is to construct orbit
libraries for varying $\rho(r)$ and match the allowed orbits to kinematics
in the form of LOSVDs---the distribution of projected velocities observed.
Some orbit libraries will contain orbits that do a good job at 
fitting the observed LOSVDs and others will not.  The best-fitting model
identifies the best-fitting $\rho(r)$.  Once we know this, we can invert
Equation (\ref{param}) to solve for $\rho_{DM}(r)$.  The trick is to vary
$\rho(r)$ in a systematic way.  This is the principal difference between our new
approach and standard Schwarzschild modeling which tries to vary $\rho(r)$
by varying the parameters that define an assumed dark matter profile.

To compute the orbit library for each model, we first calculate the
potential.   We assume axisymmetry and make use of the stellar ellipticity
to define the density at angle $\theta$ in our meridional grid.
The dark matter halo is assumed to have the same ellipticity as the stars.
We solve Poisson's equation for the potential associated with this density
distribution by decomposing $\rho(r,\theta)$ into spherical harmonics 
\citep{sio09}.  With the potential known, we launch 20,000-30,000 orbits
and integrate their motion for roughly 100 crossing times, storing position
and velocity information at each timestep.  

Orbits in axisymmetric potentials respect three isolating integrals of motion:
energy $E$, the $z$-component of angular momentum $L_z$, and the non-classical
third integral $I_3$.  By specifying all three of these quantities together,
an orbit is uniquely defined.  Unfortunately, there is no analytical form
for $I_3$ and it is generally not known a priori.  We therefore rely on the 
sampling scheme of \citet{tho04} to construct an orbit library which
uniformly samples $E$, $L_z$, and $I_3$ and thereby contains all possible
orbits for a given potential.

Each orbit in the library is given a weight $w_i$, and a set of $w_i$ are 
chosen so the observed kinematics are appropriately reproduced by the orbits
which have been weighted, averaged, and projected.
Quantitatively, we observe $N_{\mathrm{LOSVD}}$ LOSVDs in the galaxy at various
positions.  Each LOSVD contains $N_{\mathrm{vel}}$ velocity bins with 
uncertainties, so the number of observables the models must match to is
given by the product $N_{\mathrm{LOSVD}} \times N_{\mathrm{vel}}$.  The goodness-of-fit
of a model is judged by 

\begin{equation}
\chi^2 = \sum_{i=1}^{N_{\mathrm{LOSVD}}} \sum_{j=1}^{N_{\mathrm{vel}}}
\left ( \frac{\ell_{ij}^{\mathrm{obs}} - \ell_{ij}^{\mathrm{mod}}}
{\sigma_{ij}} \right )^2
\label{chi2eq}
\end{equation}

\ni
where $\ell_{ij}^{\mathrm{obs}}$ and $\ell_{ij}^{\mathrm{mod}}$ are the $j^{\mathrm{th}}$
velocity bin of the $i^{\mathrm{th}}$ LOSVD from the observations and model
respectively, and $\sigma_{ij}$ is the uncertainty in $\ell_{ij}^{\mathrm{obs}}$.

Given the freedom to choose from upwards of 10,000 orbital weights to match 
only $N_{\mathrm{LOSVD}} \times N_{\mathrm{vel}} \sim 100$ observables, a standard
$\chi^2$ minimization routine can populate the distribution function
in any number of ways that introduce unwanted noise.
To avoid distribution functions that are
noisy or unrealistic, however still consistent with the observables, we employ
a maximum entropy smoothing technique developed by \citet{ric88} and 
described in \citet{sio09}.  Instead of minimizing $\chi^2$, we maximize the
objective function

\begin{equation}
\hat{S} = -\sum_{i=1}^{N_{\mathrm{orb}}} w_i \log \left( \frac{w_i}{\Delta \Omega_i}
\right ) - \alpha_S \chi^2
\label{entropyeq}
\end{equation}

\ni
where $N_{\mathrm{orb}}$ is the number of orbits in the library, and 
$\Delta \Omega_i$ is the phase-space volume of the $i^{\mathrm{th}}$ orbit.
See \citet{sio09} for a technical description of how we calculate
phase-space volumes and maximize $\hat{S}$.

The first term in Equation (\ref{entropyeq}) is an entropy-like quantity
and the second term is $\chi^2$ from Equation (\ref{chi2eq}).  
The parameter $\alpha_S$ controls which term influences $\hat{S}$.  For small
$\alpha_S$, orbital weights are chosen to produce a smooth distribution
function at the expense of reproducing the data.  For large $\alpha_S$, 
the data are well-fit by the model ($\chi^2$ is small), but the resulting
distribution function is likely not smooth.  We determine the appropriate
$\alpha_S$ for each model using the scheme described in \citet{sio09}.  We start
with $\alpha_S=0$ and incrementally increase it until changes to $\chi^2$ 
between successive iterations are small.  Thus, the maximum entropy constraint
serves to initialize the search for the minimum when $\alpha_S=0$.  By slowly
increasing $\alpha_S$, we drive down the importance of entropy to the fit
until it no longer matters.

\begin{figure*}[t]
\centering
\includegraphics[width=15cm]{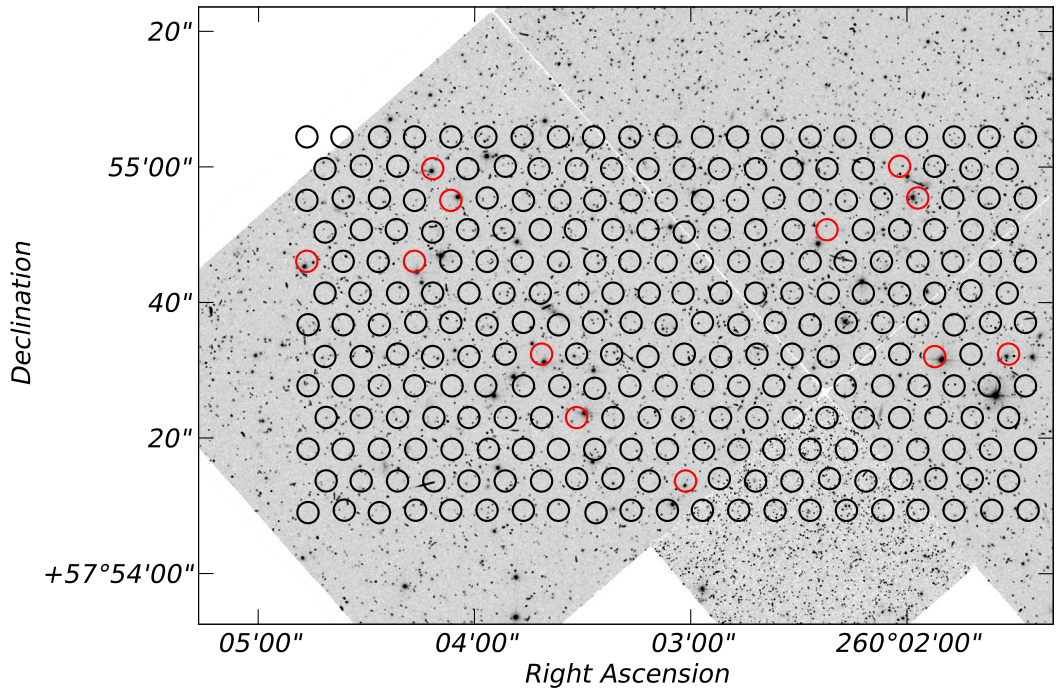}
\caption{VIRUS-W IFU overlaid on top of an \emph{HST} image from \cite{seg07}.
Red circles highlight fibers containing stars used in the determination of
the central LOSVD.  Note the \emph{HST} PSF is significantly smaller than
the typical 2\arcsec seeing at McDonald Observatory.
\label{fibmap}}
\end{figure*}

\subsection{Varying $\rho(r)$ Between Models}

The major innovation of our new modeling technique is how we choose the
density profile $\rho(r)$ of each model.  Current methods assume $\rho_{DM}(r)$
and calculate $\rho(r)$ from Equation (\ref{param}), however this requires 
knowledge of the appropriate parameterization for $\rho_{DM}(r)$.  We use a 
fundamentally different strategy and
divide $\rho(r)$ into \nbin~discrete points whose value $\rho$ at radius
$r_i$ is labeled $\rho_i$.
The \nbin~points are spaced evenly in $\log~r$ and connected by straight
line segments.  Each trial density profile is now defined by the $\rho_i$
at each of the \nbin~bins.  We run many models adjusting the values of the
$\rho_i$ so as to sample all possible density profiles.  This strategy requires
no assumptions to be made about the shape of $\rho(r)$ or $\rho_{DM}(r)$, but
it is computationally intensive for large \nbin.

The choice of \nbin~is a compromise between accuracy in reproducing 
$\rho(r)$ and computational resources.  Large values of \nbin~can make parameter
space impossibly large, while small values can be overly restrictive on
$\rho(r)$.  We have experimented with \nbin$=5$, 7, and 10.
The added freedom with \nbin$=$7 or 10 was not found to be worth the increase to
the dimensionality of parameter space.
We have also tried connecting the $\rho_i$ with splines, but found the 
additional freedom produced unrealistic density profiles.  Concern over the 
smoothness of $\rho(r)$ may be mitigated by the fact that $\rho(r)$ only
matters to our models in that it determines the potential.  As the potential
is the integral of $\rho(r)$, this introduces additional smoothness.

We extrapolate the density at the outermost
point as a power law with slope $\alpha_{\infty}$. 
The only parameters in the model are the $\rho_i$ themselves and the
extrapolation slope $\alpha_{\infty}$. The models also have the flexibility to
add a central black hole of mass $M_{\bullet}$ to the galaxy for future 
studies.

\subsection{Separating dark from stellar mass}

Once the best-fitting $\rho(r)$ is found, the task remains still to recover
the underlying dark matter density profile.  This involves finding some 
other constraint
on the stellar mass-to-light ratio.  We can often determine $M_*/L$ from 
simple stellar population (SSP) models.  The required input for SSP models
varies greatly, and different methods are appropriate depending on the galaxy
modeled.  For example, if spectra are available, stellar population synthesis
models or Lick indices can be used.  Lacking spectra, one can use the 
relations between broad-band colors and $M_*/L$ \citep{bel01}.
In nearby dSph galaxies where individual stars are resolved, color-magnitude 
diagrams can be constructed to fit for age and metallicity with isochrones. 
We can also evaluate the radial variation of $M_*/L$ as well without much
additional effort.  Spectral or photometric data need only be spatially binned
with the same procedure repeated at each bin.  
Once $M_*/L$ is calculated, stellar density is simply the product of the
(possibly radially varying) $M_*/L \times \nu(r)$.

\section{Application to Draco}

We apply our new non-parametric Schwarzschild modeling technique to study
the nearby Draco dSph.  Draco is a satellite galaxy of the Milky Way orbiting
at a distance from the sun of only $71$~kpc \citep{ode01}. At this distance
individual stars are resolved even with ground-based observatories.  
Consequently, the data we use are radial velocities of individual stars.  
Radial velocities are available for 158 stars in Draco, and we present 
radial velocities from new observations of an additional 12 stars near the 
center of Draco.

We choose Draco
because it is the most dark matter-dominated of the ``classical'' (pre-SDSS)
dSphs.  We can therefore differentiate between dark and luminous mass 
more easily since 
the baryons contribute less to the total density profile than they do in
larger galaxies.  Consequently, we can absorb larger uncertainties in $M_*/L$.
The primary science goal of this work, and
a future study of all dSphs, is to determine the functional form of the 
dark matter profile in dSphs and compare results to theoretical predictions 
by CDM. 

\subsection{Data}

\subsubsection{Kinematics}

\begin{figure}[t]
\begin{flushleft}
\includegraphics[width=10cm]{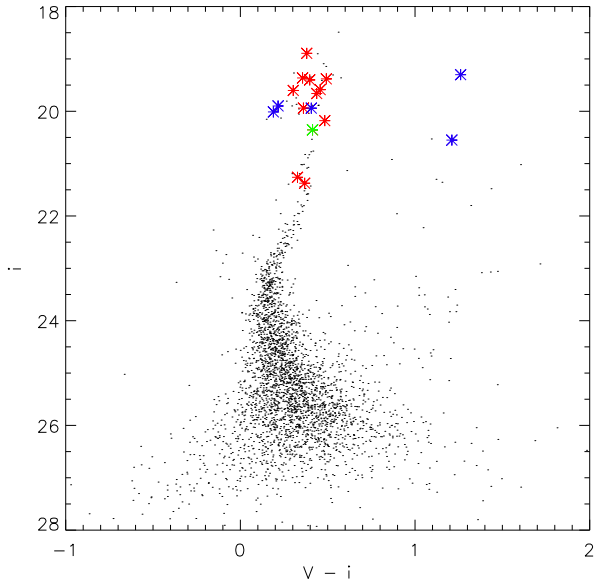}
\caption{Color-magnitude diagram of stars near the center of Draco.  
Colored asterisks are stars we observe, coded according to
their offset from Draco's systemic velocity $V_{sys}$.  Red stars have 
$|V - V_{sys}| < 30$~\kms, blue stars have $|V - V_{sys}| > 50$~\kms,
and the green star has a radial velocity between $30$ and $50$~\kms of
$V_{sys}$.
\label{cmd}}
\end{flushleft}
\end{figure}

We use a combination of published radial velocities and new observations 
for kinematics in Draco.  Data exist at larger radii 
for 158 stars \citep{kle02}, but we wish to explore the central region of
Draco in order to have the best constraint on the inner slope of $\rho_{DM}(r)$.

To accomplish this, we observe the center of Draco with the VIRUS-W integral
field unit (IFU) spectrograph \citep{fab08}
on the 2.7m Harlan J. Smith telescope at 
McDonald Observatory .  This instrument allows for a high density of stars
to be observed simultaneously, but with the drawback that fibers are not
positionable. 
There are 267 fibers that cover the
$105\arcsec \times 55\arcsec$ field of view with a $1/3$ fill factor.
We observed the spectral region covering $4900$\AA~to $5500$\AA~with a 
resolving power $R\sim 9000$.

The observations took place over the first half of 5 nights from 
2011 August 1-5 in excellent conditions.  Seeing was typically 2\arcsec or
better,
which is smaller than the 3\farcs2 diameter fibers.  The standard battery
of bias, Hg-Ne arc lamp, and twilight calibration frames were taken at the 
start of each night.  We use an early implementation of the Cure data reduction
software.  Cure is being developed as the pipeline for the Hobby-Eberly 
Dark Energy Experiment (HETDEX) \citep{hil06}.  We briefly describe steps 
taken to reduce the VIRUS-W data.  A detailed description of Cure is 
beyond the scope of this paper.  

We perform standard CCD processing steps, using the fitstools package
(described in \citealt{gos02}), to create
master bias, twilight flat, and arc lamp images for each night.  We use twilight
flats in combination with arc lamp images to determine the distortion
solution--a two-dimensional map which translates the (x,y) position of a pixel
on the CCD to a fiber number and wavelength.

Our science frames consist of 15-minute integrations of a single pointing
of the central part of the galaxy.  Prior to observing, we determined the 
optimal position of the IFU by examining \emph{Hubble Space Telescope (HST)} 
photometry of the central region \citep{seg07}.
With accurate fiber and star positions, we determined a pointing that 
maximizes the number of bright stars on fibers (see Figure \ref{fibmap}).  
There are 57 science frames with this pointing, totaling roughly 14 hours of 
integration.

We apply each night's distortion solution to the science frames 
yielding rectified, wavelength-calibrated frames.  We then collapse and 
median-combine these science frames along with the twilight flat frames.
Each night's stacked science frame is divided by the 
appropriate master flat for that night.

Since the majority of the 267 fibers in the IFU are on empty sky, we are
able to calculate an accurate sky model directly from each science frame.  
We compute this sky model for each fiber on the chip using a moving-window 
average of 20 nearby fibers.  We subtract the sky model from each frame,
and the resulting sky-subtracted frames for each night are median-combined.

\begin{figure}[t]
\begin{centering}
\includegraphics[width=8cm]{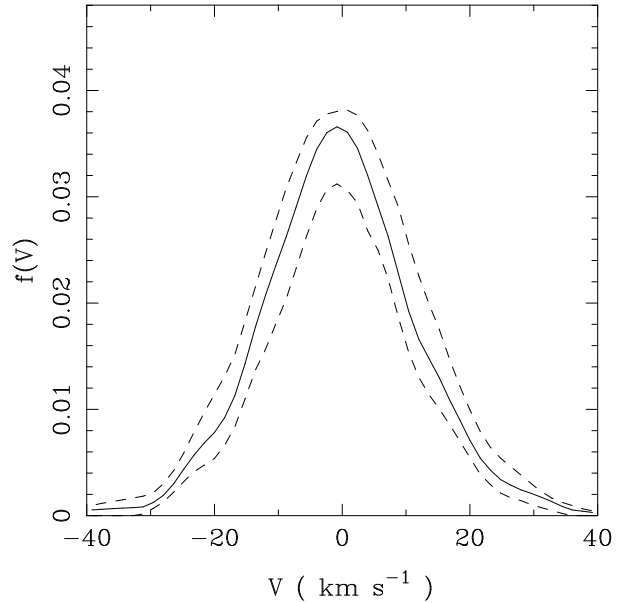}
\caption{LOSVD generated from the discrete velocities of 29 stars
\label{losvd}}
\end{centering}
\end{figure}

We extract 1-D spectra from 17 fibers containing stars.  Star 2 in our sample
is used as a velocity standard since it is the brightest member star with known 
radial velocity from \citet{arm95}.  We cross-correlate the other 16 spectra
to star 2 using the IRAF task \texttt{FXCORR}.  By cross-correlating to the
spectrum of a star with known heliocentric radial velocity in Draco, we 
automatically remove the contribution from Earth-Sun motion.  We perform
the cross-correlation analysis on the combined image, and in doing so introduce
a small bias due to the change in the heliocentric velocity correction over
the course of the observing run.  However, the magnitude of this change is
only $0.1$~\kms, much smaller than our uncertainties.  

\begin{deluxetable*}{llllll}
\tablecaption{Radial Velocities Obtained with VIRUS-W}
\tablewidth{0pt}
\tablehead{
   \colhead{Star} & \colhead{RA} & \colhead{Dec} & \colhead{$V_{\mathrm{helio}}$ \kms} &
        \colhead{$\Delta V_{\mathrm{helio}}$ \kms} & \colhead{$R_{TD}$}}

\startdata
1 & 17$^\text{h}$20$^\text{m}$14$^\text{s}$.76 & +57$^{\circ}$54\arcmin32\farcs40 & -288.1 & 2.57 & 4.76 \\ 
2 & 17$^\text{h}$20$^\text{m}$07$^\text{s}$.49 & +57$^{\circ}$54\arcmin32\farcs04 & -299.1\tablenotemark{1} & 
1.89\tablenotemark{1} & ... \\ 
3 & 17$^\text{h}$20$^\text{m}$06$^\text{s}$.12 & +57$^{\circ}$54\arcmin32\farcs40 & -293.1 & 3.99 & 8.75 \\ 
4 & 17$^\text{h}$20$^\text{m}$14$^\text{s}$.11 & +57$^{\circ}$54\arcmin23\farcs04 & -310.9 & 3.35 & 6.01 \\ 
5 & 17$^\text{h}$20$^\text{m}$12$^\text{s}$.10 & +57$^{\circ}$54\arcmin13\farcs68 & -270.6 & 3.37 & 7.47 \\ 
6 & 17$^\text{h}$20$^\text{m}$16$^\text{s}$.78 & +57$^{\circ}$54\arcmin59\farcs76 & -276.2 & 1.91 & 12.98 \\ 
7 & 17$^\text{h}$20$^\text{m}$08$^\text{s}$.14 & +57$^{\circ}$55\arcmin00\farcs12 & -258.4 & 3.89 & 7.87 \\ 
8 & 17$^\text{h}$20$^\text{m}$16$^\text{s}$.44 & +57$^{\circ}$54\arcmin55\farcs08 & -293.2 & 6.05 & 8.01 \\ 
9 & 17$^\text{h}$20$^\text{m}$07$^\text{s}$.80 & +57$^{\circ}$54\arcmin55\farcs44 & -307.6 & 4.51 & 10.92 \\ 
10 & 17$^\text{h}$20$^\text{m}$09$^\text{s}$.48 & +57$^{\circ}$54\arcmin50\farcs76 & -277.7 & 3.61 & 8.02 \\ 
11 & 17$^\text{h}$20$^\text{m}$19$^\text{s}$.10 & +57$^{\circ}$54\arcmin46\farcs08 & -292.2 & 3.23 & 10.94 \\ 
12 & 17$^\text{h}$20$^\text{m}$17$^\text{s}$.11 & +57$^{\circ}$54\arcmin46\farcs08 & -277.8 & 2.17 & 14.39 \\ 

\enddata

\label{rvtab}
\tablenotetext{1}{From \citet{arm95}}
\tablecomments{Heliocentric radial velocities for the 12 member stars observed
with VIRUS-W at the center of Draco}
\end{deluxetable*}

We list the
heliocentric radial velocities and Tonry-Davis $R_{TD}$ values
determined for the 12 stars we report as members in Table \ref{rvtab}.
The Tonry-Davis value indicates the relative strength of the primary peak in
the cross-correlation function to the average \citep{ton79}.  The right 
ascension and declination given for each star in Table \ref{rvtab} indicate 
the position of the center of the VIRUS-W fiber containing that star.

To determine membership for the 17 stars, we use the photometry of \citet{seg07}
to produce a color-magnitude diagram (CMD).  Figure \ref{cmd} presents the
resulting CMD, where the colored symbols indicate observed stars.  We also 
group the stars according to their offset from Draco's systemic velocity, which
we assume is $V_{\mathrm{sys}}=-293$ \kms~\citep{arm95}.  Stars with radial
velocity offsets greater than 50 $\text{km s}^{-1}$ are classified as 
non-members,
while stars with offsets less than 30 $\text{km s}^{-1}$ are categorized as
members.  The one star with radial velocity $V-V_{\mathrm{sys}}=32.6 \pm 3.9$~\kms
(green symbol in Figure 
\ref{cmd}) is classified as a possible member.  Possible and non-members are
discarded from further analysis, leaving 12 member stars.  Note that blind 
sigma-clipping retains these same 12 stars as members.

We have individual radial velocities for stars at positions around the
galaxy, but our models want the distribution of radial velocities at 
each position---the LOSVDs.
We group the individual velocities into spatial bins and determine
the LOSVD at each bin via an adaptive kernel density estimator 
\citep{sil86,geb96}.  In velocity space, this procedure replaces each of the
$N$ discrete observations with a kernel of width $h$ and height proportional
to $N^{-1}h^{-1}$.  We use the Epanechnikov kernel (an inverted parabola)
and sum the contribution from each discrete velocity to obtain a non-parametric
representation of the LOSVD.
The $1\sigma$ uncertainties on the LOSVDs are
calculated through bootstrap resamplings of the data (i.e. sampling with 
replacement from the N velocity measurements, see \citealt{geb96,jar12}).
In Figure \ref{losvd} we show an example LOSVD.

\begin{figure}[t]
\begin{centering}
\includegraphics[width=8cm]{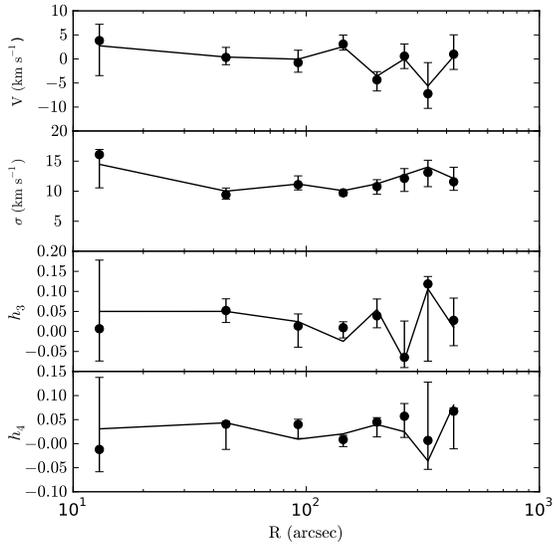}
\caption{Gauss-Hermite moments fit to the 8 LOSVDs generated from 170 radial
velocities.  The solid line is the result of our best-fit model.
\label{kin}}
\end{centering}
\end{figure}

We combine the new VIRUS-W data with 158 additional radial velocities
from the literature \citep{kle02}.  We divide these 170 radial 
velocities into 8 radial bins of roughly 20 stars each.  LOSVDs are calculated
for each of these bins,  yielding kinematics coverage over the radial
range $25\arcsec$--$1500\arcsec$ ($8\text{~pc}$--$500\text{~pc}$).
We fit Gauss-Hermite moments to the 8 LOSVDs and plot the kinematics in Figure
\ref{kin}.  This is only done for comparison purposes as the models fit
directly to the LOSVDs.  We compare the velocity dispersion as determined from
the Gauss-Hermite fit with the standard deviation of the individual velocities
(using the biweight scale; see \citealt{bee90}) in order to determine the best
value for the smoothing width $h$.  

The issue of foreground contamination frequently comes up in the study 
of dSphs using individual radial velocities.  There is always the possibility
that some fraction of the observed stars are members of the Milky Way.  These
stars would be velocity outliers and therefore artificially increase the 
measured velocity dispersion or, in our case, the width of the LOSVD.
Fortunately, the foreground Milky Way stars are well-separated in velocity
space from the \citet{kle02} sample.  Contaminants are also unlikely to have
colors and magnitudes that place them on the red giant branch of Draco's 
color-magnitude diagram.  \citet{lok05} use these two constraints to estimate
that there are of order 1-2 Milky Way contaminants in the entire
\citet{kle02} data set.


\subsubsection{Photometry}

Our models are required to not only match the observed LOSVDs but also the
three-dimensional luminosity density profile $\nu(r)$.  The first step in
obtaining $\nu(r)$ is to measure the two-dimensional surface brightness
profile.
We use the photometry of \citet{seg07} who publish a number density profile
of stars in Draco.  This profile covers the radial range from $15$\arcsec - 
$2400$\arcsec.  We extrapolate the profile as a power law out to 
$R=6000\arcsec$ by 
fitting a constant slope to the profile in logarithmic space.  To convert the
number density profile to an effective surface brightness profile, we apply
an arbitrary zeropoint shift in log space until the luminosity obtained 
by integrating the surface brightness profile is consistent with the
observed luminosity \citep{mat98}.  We plot this surface brightness profile
in Figure \ref{surf}.

We deproject the surface brightness
profile according to the procedure detailed in \citet{geb96}.  We assume
surfaces of constant luminosity density $\nu$ are coaxial spheroids and
perform an Abel inversion.  For Draco we adopt an ellipticity of $e=0.3$ 
\citep{ode01}.  We assume an inclination of $i=90^{\circ}$ for simplicity.
Inclination is typically one of the more difficult quantities to constrain
\citep{tho07}.  In addition to simplifying our models, assuming 
$i=90^{\circ}$ provides the advantage that the deprojection is unique.
For a detailed discussion of how uncertainties in viewing angle and 
geometry propagate through our models see \citet{tho07b}.

The resulting luminosity density profile we calculate has an
average logarithmic slope $\langle d\ln \nu / d\ln r\rangle  = -0.4$
inside 50~pc.  In Figure \ref{surf} we plot $\nu(r)$ and
also illustrate the positions of our kinematics data.

\subsection{Models}

Our non-parametric models of Draco use \nbin=5 radial bins spaced equally
in $\log r$ from 15\arcsec to 2000\arcsec.  We initialize our search for the
minimum with a brute force method, constructing a coarse grid in \nbin$+1$ 
dimensions from which we calculate all possible permutations of the 
\nbin~parameters and the extrapolated slope $\alpha_{\infty}$.  Additionally,
we require the density profile of each model to be monotonically decreasing
or constant.  This is a natural constraint, and it significantly lowers the
number of models needed to sample parameter space.

Once the models defining the coarse grid are evaluated, we employ an iterative
sampling scheme to focus in on and define the minimum in better detail.  
This method takes all the models with $\chi^2$ within 
$\chi^2_{\mathrm{lim}}$ of the minimum $\chi^2_{\mathrm{min}}$ as starting points.
For each starting point, a fractional step of size $\delta_i$ is taken
above and below the initial value, one at a time, for all the density bins.
If there is no change to $\chi^2_{\mathrm{min}}$, then $\delta_i$ is decreased.
This procedure is repeated until $\delta_i$ is less than a specified threshold.
Additional models are also run as needed to fill in regions of parameter space
that appear interesting.  

We do not attempt to fit
for $\alpha_{\infty}$ as we clearly do not have kinematics in that radial
range to constrain the mass.  Instead, we treat $\alpha_{\infty}$ as a 
nuisance parameter and marginalize over it in our analysis.  We restrict 
the value of the extrapolated slope to 
$\alpha_{\infty} \in \{-2, -3, -4\}$ and every $\rho(r)$ we 
sample has been run with each of these values.  These slopes are 
representative of the isothermal, NFW, and \citet{her90} density profiles.

Since Draco orbits within the dark matter halo of the Milky Way, it is
probable that is has been tidally stripped at large radii.  To account 
for this, the density is truncated at the tidal radius $r_t$ defined by

\begin{equation}
r_t \sim \left( \frac{m}{3M} \right )^{1/3} D.
\label{rt}
\end{equation}

\ni
For reasonable values of the Milky Way's mass $M$, Draco's mass $m$, and the
Galactocentric radius of Draco's orbit $D$ (assumed circular), 
Equation \ref{rt} gives an approximate tidal radius $r_t \approx 3$~kpc.
We therefore truncate $\rho(r)$ at this radius.  We also assume the dark 
halo in Draco
has the same flattening as the stars and therefore leave $q_{DM}$ fixed at
0.7.  In the future we plan to investigate models with varying $q_{DM}$, 
however that is not the focus of this paper.



\section{Results}

\begin{figure*}[t]
\begin{centering}
\includegraphics[width=15cm,angle=0]{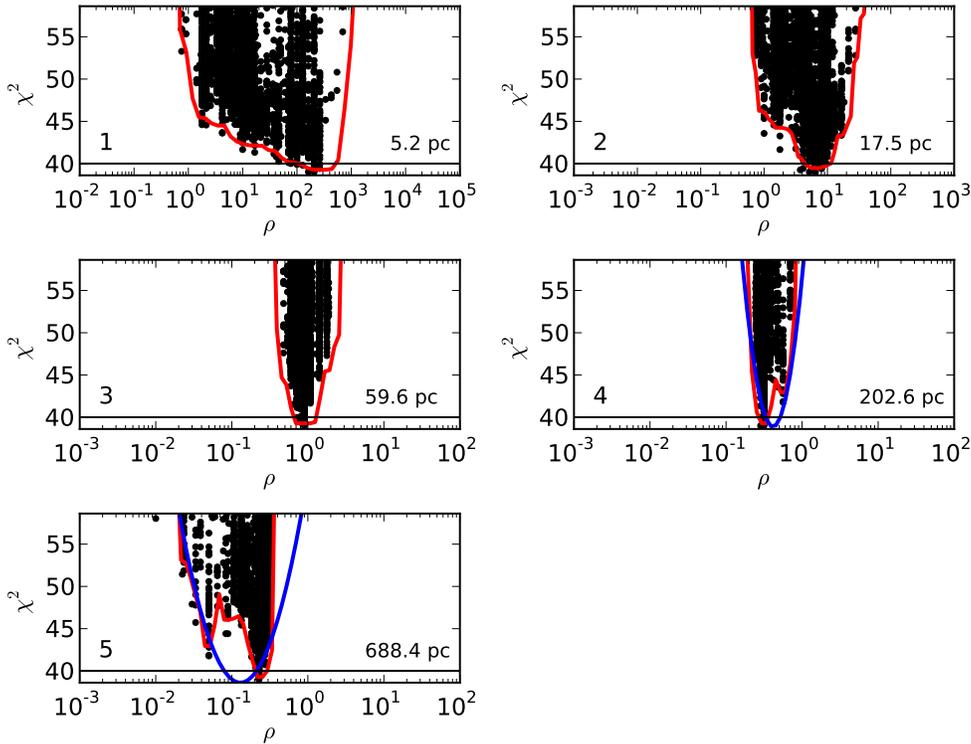}
\caption{$\chi^2$ curves for all of the $\rho_i$ parameters.  Each black
dot represents a single model (combination of $\rho_1, \rho_2, \ldots \rho_5$)
and the red curve is a smoothed fit to the minimum.  The red curve in any panel
therefore is the $\chi^2$ curve marginalized over the other density points.
The unit of density is $M_{\odot}$~pc$^{-3}$.  In panels 4 and 5, the blue 
curve is a parabola in $\log \rho$ that we use to interpolate between
two local minima.
\label{chi2}}
\end{centering}
\end{figure*}

The $\chi^2$ curves for all the $\rho_i$ are plotted in Figure \ref{chi2}.
Each dot represents a single model, and the red curve is a smoothed fit to the 
minimum.  We obtain the red curve through a smoothing process that is similar
to a boxcar average, except that we take the biweight of the 7 lowest $\chi^2$
values within the boxcar.  This method is therefore less sensitive to outliers
than a traditional boxcar average.  When determining a smoothed fit to the 
minimum, one is tempted to use only the 
points with the lowest $\chi^2$.  However, numerical noise causes models to 
scatter to both higher and lower $\chi^2$ in some bins.  
This is difficult to see by eye because scatter to higher values of $\chi^2$
causes the models to blend in with the black points in Figure \ref{chi2}
while scatter to lower $\chi^2$ makes models appear to stand out.  The 
sliding biweight robustly picks out the center of this distribution 

\begin{figure}[t]
\begin{centering}
\includegraphics[width=8cm]{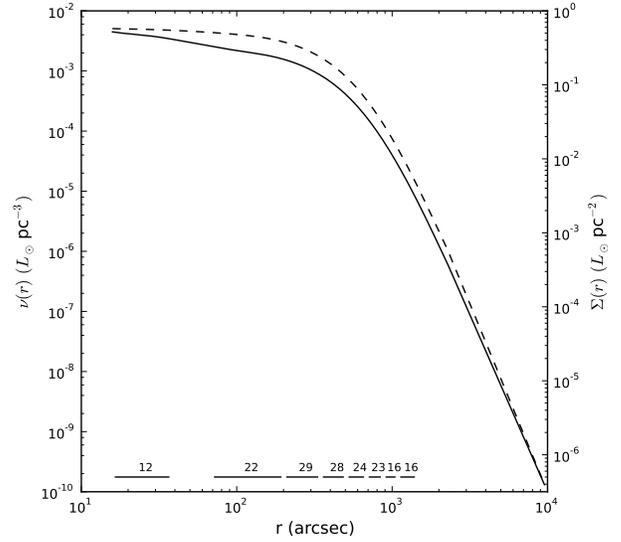}
\caption{Surface brighness profile $\Sigma (r)$ (dashed) and deprojected 
luminosity density profile $\nu (r)$ (solid) used in our models.  Horizontal
lines near the x-axis indicate the radial position of our kinematics bins.
Numbers refer to the number of radial velocities used per bin.  Note the 
central location of the new VIRUS-W data (innermost bin) in comparison 
to existing data.
\label{surf}}
\end{centering}
\end{figure}

The red curve plots $\chi^2(\rho_i)$ for each radial bin, and therefore
gives an indication of the model-preferred density at radius $r_i$.  
We estimate the $1\sigma$ uncertainties on 
each of the $\rho_i$ by determining the portion of each parameter's 
$\chi^2$ curve, marginalized over all other parameters,
that lies within $\Delta \chi^2=1$ of the overall minimum. 
Figure \ref{chi2} shows this limit as a horizontal line whose intersection
with the red curve indicates the $1\sigma$ range of the density at bin $i$.
In all further analysis we identify the midpoint of this range as the
best-fitting value and report uncertainties as symmetric about this value.

In two cases, bins 4 and 5, there are secondary minima that extend almost to 
the $\Delta \chi^2=1$ line but not quite.  It is likely that
with perfect coverage of parameter space the area between these minima would
be filled in.  However, available computational resources limit the extend to
which we can sample parameter space.  In order to be more conservative in our 
analysis we fit a quadratic in $\log \rho$ to these minima, centered roughly
on the midpoint between them (blue curves in Figure \ref{chi2}).

The best-fitting model has unreduced $\chi^2_{\mathrm{min}}=9.1$, and the number
of observables our models fit to is 
$\nu = N_{\mathrm{LOSVD}} \times N_{\mathrm{vel}}= 8 \times 15 = 120$.  If we were to
naively calculate a reduced $\chi^2$, we would estimate $\chi^2_{\nu}=0.08$.
This low value of $\chi^2_{\nu}$ results from an overestimation of the number
of independent degrees of freedom $\nu$.  The adaptive kernel density 
estimator we use to compute the LOSVDs introduces correlations among
neighboring velocity bins, therefore reducing the number of truly independent
degrees of freedom.  

To account for this, we consider the Gauss-Hermite parameterizations
of our best-fitting model (solid line in Figure \ref{kin}) and input
LOSVDs (points with error bars in Figure \ref{kin}).  This model has 
$\chi^2_{{\nu}_{GH}}=0.33$ where $\nu_{GH}$ is 4 Gauss-Hermite parameters $\times$
8 LOSVDs$=32$.  This $\chi^2_{{\nu}_{GH}}$ is still less than 1, however it is 
more consistent with previous studies \citep{geb03} and may be due to 
correlations among the Gauss-Hermite parameters (e.g. \citealt{hou06}).
We use $\chi^2_{{\nu}_{GH}}$ to calculate the appropriate scaling to apply to our
models which use the LOSVDs in determining $\chi^2$.  We scale all un-reduced
$\chi^2$ values by the factor $\chi^2_{{\nu}_{GH}} / \chi^2_{\nu}=4.3$


\subsection{Obtaining $M_{*}/L$}

We have so far identified the best-fitting total density profile.  In order to
study the dark matter profile we must subtract the stellar density profile
$\rho_*(r)$.  This involves finding an independent constraint on the stellar
mass-to-light ratio $M_*/L$.  
Using stars within the central 5\arcmin~of Draco, we construct a 
$\mathrm{g}\arcmin - \mathrm{i}\arcmin$ color-magnitude diagram (CMD) from the 
photometry of \citet{seg07}.  We fit isochrones \citep{mar08} to the CMD, 
corrected for Galactic extinction \citep{sch98}, so that we may 
determine the age and metallicity of the stellar population.

\begin{figure}[t]
\begin{centering}
\includegraphics[width=9cm]{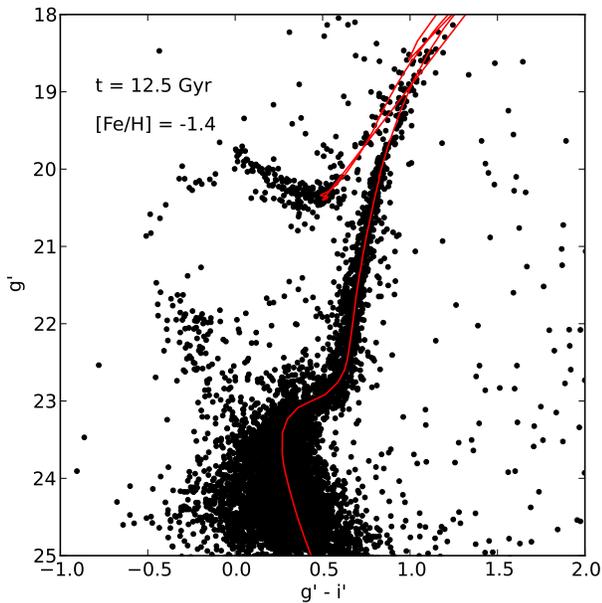}
\caption{Color-magnitude diagram of stars within the central 5\arcmin of
Draco. From left to right, we plot isochrones of 
$(t_{age} \times 10^9 \text{yr}, 
\mathrm{[Fe/H]})$ = $(11.5, -1.6)$, $(12.5, -1.4)$, and $(13.5, -1.3)$.  
The solid red line is the $(12.5, -1.4)$ isochrone we use when determining 
$M_*/L_V$.
\label{iso}}
\end{centering}
\end{figure}

Figure \ref{iso} shows the CMD with our best isochrone fit.  The red giant
branch is well-defined, and we obtain a sensible fit with age 
$t_{\mathrm{age}}=12.7$~Gyr and metallicity $[\mathrm{Fe/H}]=-1.4$. 
Using the SSP models from \citet{mar05} we are able to convert 
$t_{\mathrm{age}}$ and $[\mathrm{Fe/H}]$ to a V-band stellar mass-to-light
ratio $M_*/L_V=2.9 \pm 0.6$.  Uncertainties in $M_*/L_V$ represent the 
spread in SSP predictions when different initial mass functions
are assumed in the models.


\subsection{The Dark Matter Profile}

\begin{figure}[t]
\begin{centering}
\includegraphics[width=9cm]{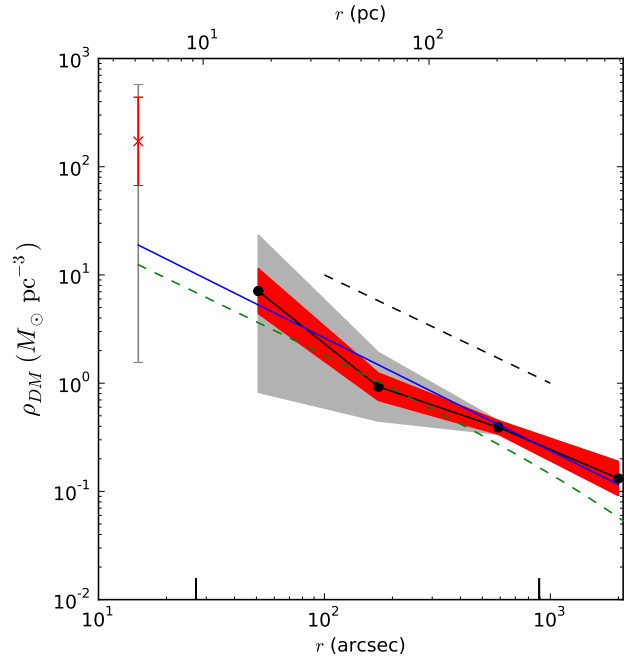}
\caption{Best-fitting dark matter density profile in Draco.  The red shaded 
region
represents the point-wise 68\% confidence band for $\rho_{DM}(r)$ 
($\Delta \chi^2=1$), with the
solid black line derived from forcing symmetric logarithmic errors.  The gray
shaded region is the 68\% confidence band on $\rho_{DM}(r)$ considering all
parameters jointly ($\Delta \chi^2=7.04$).  We plot the innermost point
(excluded from all futher analysis) an an error bar with the same color
scheme.  The
solid blue line is the best power law fit to the profile, and the dashed line
shows an $r^{-1}$ NFW-like profile.  We plot the best-fitting NFW halo from a 
small grid of parametric models as the dashed green line.
Vertical lines along the x-axis indicate the radial range of our kinematic
data.
\label{dmdens}}
\end{centering}
\end{figure}

With $M_*/L_V$ determined from stellar population models, we can subtract
$\rho_*(r)$ from the best-fitting total density profile obtained during the
modeling procedure.  We plot the resulting dark matter profile in Figure 
\ref{dmdens}.
The red band is the 68\% confidence band for each density point,
marginalized over the others, and the gray band shows the 68\% confidence
band of all the parameters jointly (at $\Delta \chi^2=7.04$).  

From Figure \ref{dmdens}, it seems plausible that $\rho_{DM}(r)$ can be fit by a 
power law of the form $\log \rho_{DM} = \alpha \log r + \beta$ with the 
exception of perhaps the innermost data point.  The slope
of this fit $\alpha$ can be directly compared to both theoretical predictions
and observations of similar dSphs.  The innermost
point, however, is puzzling.  Its value indicates a large central density
and a departure from the power-law nature of the outer profile.
Further puzzling is that its point-wise uncertainty (plotted as a red error bar)
indicates strong 
constraint despite the fact that we have no kinematic data in this region
of the galaxy.  We speculate that, in the absence of such data, models are 
able to arbitrarily increase the central density.  Since the volume of this
inner bin is small, the total amount of mass added is negligible.  
With no kinematics in this 
region, models can easily absorb this mass without affecting $\chi^2$.   
We therefore exclude the innermost point in all further analysis.

The resulting power law fit to the outer four points is shown in 
blue in Figure \ref{dmdens}.  We characterize the uncertainty in this fit 
by constructing 1,000 Monte Carlo realizations with noise added to the density
profile.  We draw each point
point $i$ randomly from a Gaussian distribution with mean $\log \rho_i$ and
dispersion given by the width of the $1\sigma$ confidence band at point $i$
in Figure \ref{dmdens}.  We repeat the fit for each realization, and
determine the $1\sigma$ uncertainties on $\alpha$ from the 68\% span of
this distribution.  This procedure yields $\alpha = -1.0 \pm 0.2$.  
None of the 1,000
realizations has a slope $\alpha > -0.45$ strongly indicating that the
galaxy is not cored for $r \gsim 20$~pc.

\subsection{Orbit Structure}

\begin{figure}[t]
\begin{centering}
\includegraphics[width=9cm]{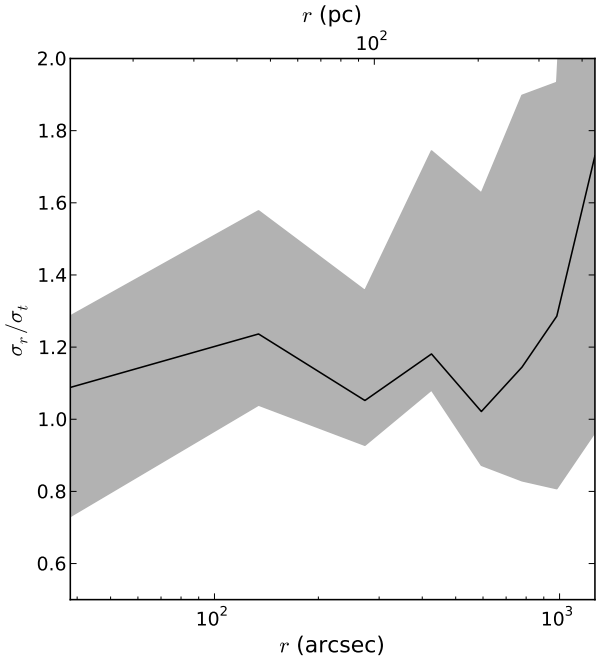}
\caption{Ratio of the radial to tangential components of the velocity 
dispersion.  Values of $\sigma_r/\sigma_t$ different from unity indicate
anisotropy.  The black line is our best-fitting model.
\label{aniso}}
\end{centering}
\end{figure}

Once we have determined the best-fitting model, we can calculate the 
internal (unprojected) moments of the distribution function at each of the
bins in our meridional grid.   Of interest is
the anisotropy in the velocity dispersion tensor, which we quantify with the
ratio $\sigma_r/\sigma_t$---the ratio of radial to tangential
anisotropy in the galaxy.  We define the tangential anisotropy $\sigma_t$ as

\begin{equation}
\sigma_t \equiv \sqrt{ \frac{1}{2}( \sigma^2_{\theta} + \sigma^2_{\phi} +
  v^2_{\phi} ) }
\label{vteq}
\end{equation}

\noindent
in spherical polar coordinates
where $v_{\phi}$ is the rotational velocity.  Streaming motions in the $r$ and
$\theta$ directions are assumed to be zero.  We plot $\sigma_r/\sigma_t$ in
Figure \ref{aniso}.  Since the LOSVDs we use in Draco contain contributions
from stars 
at all angles $\theta$, we average $\sigma_r$ and all quantities in 
Equation (\ref{vteq}) when calculating $\sigma_r/\sigma_t$.  Consequently,
we lose the ability to evaluate anisotropy as a function of $\theta$.
This can be avoided if better kinematics coverage is available,
either through more stars with radial velocities in dSphs or 
two-dimensional integral-field spectroscopy in more distant
galaxies.  Fortunately most other large dSphs in the Local Group have
many more radial velocities publicly available.

We plot $\sigma_r/\sigma_t$ in Figure \ref{aniso} over the radial range that our
LOSVDs sample.  
We determine the uncertainties in $\sigma_r/\sigma_t$ by the maximum/minimum
values of $\sigma_r/\sigma_t$ for models within $\Delta \chi^2=7.04$ of
$\chi^2_{\mathrm{min}}$ ($1\sigma$ for \nbin$+1$ degrees of freedom).We find 
evidence for radial anisotropy at all radii, 
consistent with the ``tidal stirring'' theory describing the origin of
the Milky Way dSphs \citep{lok10,kaz11}.
Uncertainties are large on $\sigma_r/\sigma_t$, likely due to the small number
of radial velocities available as kinematic constraint.  To constrain the 
anisotropy better, more radial velocities are needed. 

\section{Discussion}

\subsection{Improvement over Parametric Methods}

Since we eventually fit our non-parametric dark matter profile with a power law,
one can ask why we do not initially use a power law-parameterized profile.
This would seem advantageous, especially given the large parameter space 
required by non-parametric methods.  This reasoning, however, relies on the 
assumption that we know the profile is a power law a priori.  The point of this
study is to relax this assumption and see what type of profile comes out of the 
modeling, rather than impose unjustified interpretation on the problem.  It 
happens that Draco hosts a nearly power law density profile, but by not 
assuming this a priori we allow more general models to be explored.  As a rough
check that our models have converged to a global minimum, we run a small grid 
of parametric models with an NFW dark matter density profile.  The best-fitting
of these models is plotted in green in Figure \ref{dmdens}.

\subsection{Interpreting the Dark Matter Profile}
\label{discuss}

It is important to note that we only constrain the dark matter density profile 
over little more than a decade in radius from $20-700$~pc.  One could easily
imagine our power law fit changing from $\alpha=-1$ to a core ($\alpha=0$)
inside of $r \sim 20$~pc.  Likewise, the slope may also change at larger radii
than $r \sim 700$~pc without our knowledge.  The NFW density profile has an 
outer slope $\alpha = -3$ for $r \gg r_s$, but our profile does not change slope
within our model grid.  This could indicate that $r_s \gg 700$~pc, but without
knowledge of the outer slope we cannot say with certainty that the profile is 
NFW-like.  

Recent cosmological $N$-body simulations have
been found to produce density profiles shallower than the traditional
$\alpha = -1$ cusps \citep{sta09,nav10}.  Many authors suggest that dark matter
profiles are best parameterized by the Einasto profile 
\citep{nav04,mer05,gao08,nav10} where the slope
varies with radius according to a power law $\alpha (r) \propto r^n$.
These profiles can have shallower cusps than NFW, but do not have constant
slopes over a large range in radius.  Our non-parametric density profile is
well-fit by a single power law from $20 \lsim r \lsim 700$~pc, but, again,
this is a fairly narrow range in radius.  Our models cannot rule out an 
Einasto-like change 
in slope outside this radial range.  More kinematics are needed to characterize
the density profile at large and small radii.

When calculating the potential, we allow the outer slope of $\rho(r)$ 
to vary between $2 \leq \alpha_{\infty} \leq 4$ for $r>700$~pc, but, 
unsurprisingly, we are unable to constrain $\alpha_{\infty}$.
Tidal effects may also alter the shape of $\rho_{DM}(r)$ since Draco is orbiting
within the dark matter halo of the Milky Way.  The tidal radius calculated
from Equation (\ref{rt}) is sufficiently
large that tides are unlikely to affect the stellar component, but 
$\rho_{DM}(r)$ at large radii could be affected.  If this is the case, 
$\rho_{DM}(r)$ would decline more steeply than expected and the total mass 
enclosed would be smaller than what we calculate.

The cuspy $\alpha=-1$ dark matter profile we find in Draco stands in contrast
to many other observational studies of dSphs that find $\alpha=0$ cores
\citep{gil07,wal11,jar12}.  The effects of baryons are still not 
well-understood, and could potentially drive $\alpha$ to different values
on a galaxy-by-galaxy basis.  These effects are the sum of at least two
competing processes.  Adiabatic compression \citep{blu86} draws in dark matter
boosting the central $\rho_{DM}$ and driving $\alpha$ to more negative values.
On the other hand, feedback from star formation and supernovae can cause
strong outflows \citep{nav96b, bin01} which can in turn remove dark matter
from the centers of galaxies, reshaping cuspy profiles into $\alpha=0$ 
cores.  

In a recent paper, \citet{gov12} use high resolution cosmological
$N$-body simulations with a fully hydrodynamical treatment of baryons to
test these two competing effects in low-mass dwarf galaxies.  They find that
the cuspiness of the dark matter halo is directly related to the amount
of star formation activity in the galaxy.  This is expressed as a correlation
between $\alpha$ and stellar mass $M_*$.  Their interpretation is that
galaxies that form more stars (larger $M_*$) have more supernovae and a 
greater potential to turn a cuspy dark matter profile into a core.  Using their
least-squares fit to the $M_*$-$\alpha$ correlation, they predict 
$\alpha \approx -1.3$ (at $500$~pc) for Draco's stellar mass.  This is in 
approximate agreement with our measured value of $\alpha=-1$.

Perhaps owing to the lack of stellar velocities available in Draco
compared to other dSphs, there are not many studies investigating its
dark matter profile through dynamical models.  A rough comparison can be made 
with \citet{lok05} who fit profiles with an inner slope
of $\alpha=-1$ and an outer exponential cutoff at large radii.  They find
a total mass-to-light ratio that varies with radius between 
$100 \gsim M_{\mathrm{tot}}/L_V \gsim 1000$~in the inner $\sim 700$~pc.
These values are comparable to the total mass-to-light ratio 
we calculate in the inner $\sim 300$~pc.  However, unlike 
\citet{lok05} we do not impose an exponential cutoff in $\rho_{DM}(r)$ at large
radii.  Our calculated $M_{\mathrm{tot}}/L_V$ therefore rises sharply at large
radii where the stellar luminosity profile is decreasing much faster than 
$\rho_{DM}(r)$.  

Importantly, $M_{\mathrm{tot}}/L_V \gg M_*/L_V=2.9\pm 0.6$ (the stellar 
mass-to-light ratio we derive from SSP models) at all radii.  This means 
we can confidently state that Draco is dark matter-dominated
at all radii, allowing us to easily absorb errors in $M_*/L_V$ from SSP
models.  In other words, when determining $\rho_{DM}(r)$ from Equation (1)
the uncertainty in $\rho(r)$ dominates the uncertainty in stellar density
since the product $M_*/L \times \nu(r)$ is much smaller than $\rho(r)$.
This is one of the reasons we choose to test this non-parametric 
technique on Draco first.  In the future we plan to extend this analysis
to the remaining Local Group dSphs, which are also thought to be dark 
matter-dominated everywhere.

\subsection{Draco's Mass}
\label{masssec}

\begin{figure}[t]
\begin{centering}
\includegraphics[width=9cm]{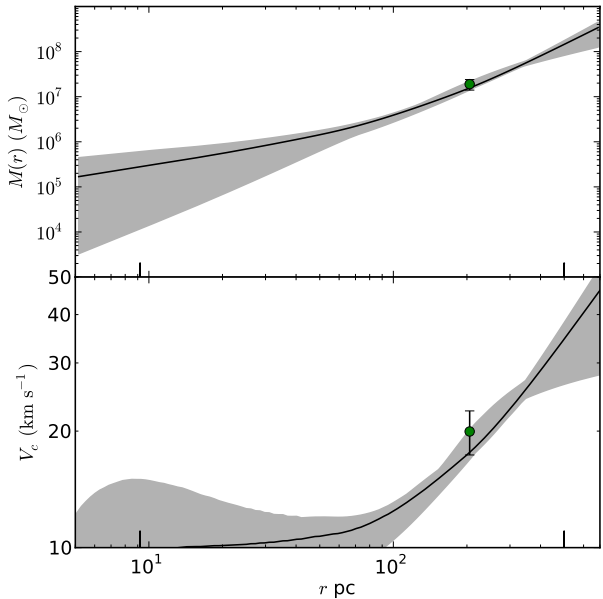}
\caption{(Top): Enclosed Mass profile of our best-fitting model (black line)
and $1\sigma$ confidence region.  The green point is the \citet{wol10}
mass estimator.  (Bottom):  Circular speed profile and $1\sigma$ confidence
region.  Colors are the same as above. Vertical tick marks on the x-axis 
represent the range of our kinematics coverage.
\label{mr}}
\end{centering}
\end{figure}

We plot the enclosed mass profile of our models in Figure \ref{mr}.  The 
shaded region is the $1\sigma$ confidence band derived from the extreme
values of $M(r)$ for all models within $\Delta \chi^2=5.84$ of the minimum
($1\sigma$ for \nbin$=5$ free parameters, marginalizing over $\alpha_{\infty}$).
The vertical ticks on the x-axis represent the radial extent of our 
kinematics coverage.  
From this plot it is apparent that, despite its low luminosity and stellar mass,
Draco lives in a dark matter halo that is surprisingly massive.  

An interesting
comparison can be made with the brightest dSph Fornax, roughly two orders of
magnitude higher in luminosity.  If we compare the mass enclosed within a
common physical radius of 300~pc, we find that for Draco
$M_{300}\equiv M(r=300\text{~pc})=3.8^{+0.84}_{-0.29} \times 10^7 \, M_{\odot}$, 
and \citet{jar12} measure 
$M_{300}=3.5^{+0.77}_{-0.11} \times 10^6 \, M_{\odot}$ for Fornax.  Of course,
Fornax is much more extended than Draco so it is sensible to also compare
the mass enclosed within the deprojected half-light radius of each galaxy's
stellar component.  For Draco we measure 
$M_{1/2}\equiv M(r=r_e)=1.6^{+0.6}_{-0.2} \times 10^7 \, M_{\odot}$, and in 
Fornax \citet{jar12} measure $M_{1/2}=5.8^{+1.0}_{-0.2} \times 10^7 \, M_{\odot}$.
We would prefer to compare the total mass of each galaxy, but there are no
kinematic tracers far enough out in the halo that the density profile declines
sharply enough to keep mass finite for any dSph.  Consequently, we cannot
constrain the total mass observationally and we must rely on comparions
to simulations (Section 5.4).


We also use our dynamical models to compare our measurement of $M_{1/2}$ with
the convenient mass estimator proposed by \citet{wol10}
(see \citealt{wal09b} and \citealt{cap06} for similar formulae).  This 
formula relates $M_{1/2}$ to the directly observable  luminosity-weighted 
line-of-sight velocity dispersion $<\sigma^2_{LOS}>$ and projected half-light
radius $R_e$.  The \citet{wol10} mass estimator is written as:

\begin{equation}
M_{1/2} \approx 4 G^{-1} R_e <\sigma^2_{LOS}> 
\label{joe}
\end{equation}

\ni 
and \citet{wol10} give a theoretical argument for why the effect of 
anisotropy is minimized near $r_e$ for a variety of stellar systems in
spherical symmetry.  

For a more fair comparison of Equation (\ref{joe}) to our models we calculate
$M_{1/2}$ from our data set, not the value listed in \citet{wol10}.
We use $<\sigma^2_{LOS}>=11.3 \pm 1.6$~\kms,
calculated directly from our data in Figure \ref{kin}, as well as 
$R_e=158.1$~pc and $r_e=205.2$~pc which we derive from the photometry in 
Figure \ref{surf}.  This calculation yields an estimated 
$M_{1/2}=(1.9 \pm 0.5) \times 10^7 \, M_{\odot}$, in excellent agreement with
the mass calculated from our models.  We plot the estimated $M_{1/2}$ as the
green point in Figure \ref{mr}.

\subsection{Comparing Draco to CDM Simulations}

We can also gain insight into the properties of Draco's dark matter halo
by examining the circular speed profile $V_c(r)$ plotted in the lower
panel of Figure 
\ref{mr} .  The green point plotted is 
$V_{1/2}=\sqrt{GM_{1/2}/r_{1/2}}= 20.0 \pm 2.6$~\kms using
our value of the \citet{wol10} mass estimator.  In a recent paper,
\citet{boy12} match the observed $V_{1/2}$ of Local Group dSphs to subhalos
around a Milky Way-like halo in the Aquarius Simulation \citep{spr08} to derive
constraints on each dSph's maximum circular speed $V_{\mathrm{max}}$---a quantity
directly related to the total halo mass.  \citet{boy12} find that
this estimate of $V_{\mathrm{max}}$ is usually $20-30$~\kms~smaller than the 
$V_{\mathrm{max}}$
they obtain through abundance matching.  These results lead them to conclude
that the Local Group dSphs are dynamically inconsistent with the types of
halos they are predicted to inhabit from abundance matching.  

We are in a position to investigate this claim directly in Draco.  We do
not need to match our $V_{1/2}$ to simulations in order to gain knowledge of
$V_c(r)$; we calculate the latter directly, and not just at the half-light
radius.  Interestingly, much of our circular speed profile lies above the
$V_{\mathrm{max}}=20.5^{+4.8}_{-3.9}$ predicted by \citet{boy12}.  At $r=500$~pc,
the radius where we run out of kinematic tracers and can therefore no longer
robustly constrain the mass, we find $V_c=34.6^{+3.5}_{-8.2}$~\kms.  We can take
the lower bound of $V_c$ here as lower limit on $V_{\mathrm{max}} \geq 26.4$.
The scaling relations between total mass and $V_{\mathrm{max}}$ for subhalos
\citep{spr08} imply a lower
limit on Draco's total mass of $M \geq 1.0 \times 10^9 \, M_{\odot}$.  

Ours is not the first study to suggest that Draco lives in a halo with such
a large mass.  \citet{pen08} demonstrate that a family of NFW halos 
with varying $V_{\mathrm{max}}$ and $r_{\mathrm{max}}$ are consistent with the 
stellar kinematics of any King model embedded in an NFW halo.  They break this
degeneracy by invoking the correlation between $V_{\mathrm{max}}$ and 
$r_{\mathrm{max}}$ found in CDM simulations (e.g. \citealt{bul01}).
Their study suggests that Draco is the most massive of the Milky Way dSphs
with $V_{\mathrm{max}} \approx 35$~\kms.

The comparison between Draco and Fornax is interesting as the two galaxies
are separated by almost two orders of magnitude in luminosity but may have
similar masses.  Since Draco's inner halo is nicely fit by the NFW density profile (Figure 
\ref{dmdens}), we can rely on simulations to extrapolate a total mass 
$M \geq 1.0 \times 10^9 \, M_{\odot}$.  However, multiple independent studies
using different methods
suggest that Fornax does not live in an NFW halo \citep{goe06,wal11,jar12},
and we therefore should not use the NFW formalism to predict a total mass from
its $V_{\mathrm{max}}$.  Still, the similarity in the galaxies' values of 
$M_{1/2}$ and $M_{300}$ suggests that
the simplest abundance matching models, which require a one-to-one mapping
between luminosity and total mass, may not appropriately describe the dSphs.
If Draco and Fornax do indeed have similar masses, despite vastly different
baryonic properties, then there must be substantial stochasticity in the 
galaxy formation process at the dSph mass scale.  Even without comparing to
Fornax, it is clear that Draco's baryonic properties do not map in the 
expected way to its halo mass.

\begin{acknowledgments}

K.G. acknowledges support from NSF-0908639.  
We thank Dave Doss and the staff at 
McDonald Observatory for observing support.  We gratefully acknowledge fruitful 
discussions with Louis Strigari, Remco van den Bosch, Glenn van de Ven, and
Andrea Maccio.  This work would not be possible
without the state-of-the-art computing facilities available at the Texas
Advanced Computing Center (TACC).


\end{acknowledgments}

\bibliographystyle{apj}

\end{document}